\newcommand{\uno}{{\bf 1}}

\newcommand{\treintaycinco}{{\bf 35}}
\newcommand{\cincuentayseis}{{\bf 56}}
\newcommand{\setenta}{{\bf 70}}

\newcommand{\cuatrocientoscinco}{{\bf 405}}
\newcommand{\setecientos}{{\bf 700}}
\newcommand{\milcientotreintaycuatro}{{\bf 1134}}

\documentclass[epj]{svjour}
\usepackage{graphics}
\begin{document}
\title{Large $N$ Weinberg-Tomozawa interaction and spin-flavor symmetry}
\author{C. Garc{\'\i}a-Recio
\and J. Nieves 
\and L.L. Salcedo 
}                     
\institute{
Departamento de F{\'\i}sica At\'omica, Molecular y
Nuclear, Universidad de Granada, E-18071 Granada, Spain
}
\date{Received: date / Revised version: date}
%
\abstract{The construction of an extended version of the
Weinberg-Tomozawa Lagrangian, in which baryons and mesons form
spin-flavor multiplets, is reviewed and some of its properties
discussed, for an arbitrary number of colors and flavors. The coefficient tables
of spin-flavor irreducible representations related by crossing
between the $s$-, $t$- and $u$-channels are explicitly constructed.
} 
\maketitle

\section{Introduction}

Spin-flavor symmetry, the symmetry by which SU($N_f$) is promoted to
SU($2N_f$), was fashionable in the early days of the quark model
\cite{Gursey:1992dc} due to its predictive
power and successes in the description of hadronic properties, but was
largely left aside in favor of QCD when the latter theory was
developed. In addition symmetries including together Poincar\'e and
internal groups were shown to be of necessity of approximated type
\cite{Coleman:1967ad}. As it turned out, large $N$ QCD (one of the few
techniques available to attack QCD in the non perturbative regime) was
shown to display spin-flavor symmetry for the lightest baryon of the
theory \cite{DJM94}.

The success of chiral unitarization schemes in the study of
meson-baryon resonances within a Bethe-Salpeter approach \cite{weise},
indicated the opportunity of considering their large $N$ versions. In
particular, in the $s$-wave $0^-$meson--$1/2^+$baryon sector,
appropriate to investigate negative parity baryons, a large
simplification of the formalism was obtained if the introduction of
spin-flavor multiplets for baryon was complemented with the similar
approach in the meson sector, thereby including vector mesons in the
game \cite{GNS05}. This naturally leads to an extended version of the
$s$-wave Weinberg-Tomozawa (WT) Lagrangian \cite{Wein-Tomo} with
spin-flavor symmetry, the construction of which is reviewed below, in
the first part of the paper. This interaction turns out to have a very
constrained form which follows entirely from chiral symmetry
considerations. Chiral symmetry consistent with spin-flavor has been
advocated by Caldi and Pagels \cite{CP76}. In the second part of the
paper we present new and detailed results concerning crossing
coefficients. They describe how interactions driven by specific
SU($2N_f$) irreducible representation in the $t$- or $u$-channels are
decomposed in the $s$-channel.

\section{WT interaction and spin-flavor symmetry}

In what follows $N$ denotes the number of colors, $N_f$ the number of
flavors and $n=2N_f$.  In the large $N$ limit, the spin-flavor SU($n$)
symmetry becomes exact for the QCD baryon \cite{DJM94}. The baryon is
described by a field ${\cal B}^{i_1\cdots i_N}(x)$. The label $i$
contains spin and flavor and runs from 1 to $n$. This field is a fully
symmetric SU($n$) tensor, i.e., falls in the representation with Young
tableau $[N]$. (The fully antisymmetric color factor of the
wavefunction is implicit.)  For $N=N_f=3$ this is the irrep
$\cincuentayseis$ of SU(6), so for general $N$ and $N_f$ it is
sometimes denoted by ``$\cincuentayseis$''.  Assuming spin-flavor
symmetry in the mesonic sector (not a direct consequence of large $N$
QCD), the lightest mesons fall in the adjoint representation of
SU$(n)$, ``$\treintaycinco$'' with tableau $[2,1^{n-2}]$, with linear
field $\Phi^i{}_j(x)$ and non linear field ${\cal U}=\exp(2i\Phi/f)$.

The baryon and linear meson fields transform and are normalized as
$Q^{i_1}\cdots Q^{i_N}$ and
$Q^i\overline{Q}_j-(1/n)Q^k\overline{Q}_k\delta^i{}_j$,
respectively, where $Q^i$ and $\overline{Q}_i$ are quark and the
antiquark fields,
\begin{equation}
Q^i\to U^i{}_jQ^j,~~
\overline{Q}_i\to (U^i{}_j)^*\overline{Q}_j=
\overline{Q}_j  (U^{-1})^j{}_i,
~~ U\in {\rm SU}(n).
\end{equation}

As in the SU(3) case, the WT Lagrangian is obtained from the free
baryonic Lagrangian by replacing the derivative by a covariant
derivative which includes the mesons
\begin{equation}
{\cal L}= \frac{1}{N!} {\cal B}^\dagger_{i_1\cdots i_N}
\left(
i\nabla_0-M+\frac{1}{2M}\vec\nabla^2
\right)
{\cal B}^{i_1\cdots i_N} \,.
\end{equation}
The covariant derivative $\nabla_\mu=\partial_\mu+A_\mu$ has connection
\begin{eqnarray}
A_\mu(x) &=& \frac{1}{2}(u^\dagger \partial_\mu u
+ u \partial_\mu u^\dagger)
\nonumber \\
&=& \frac{1}{2f^2}[\Phi,\partial_\mu\Phi]+{\cal O}(\Phi^3)
\end{eqnarray}
(where $u^2={\cal U}$). This yields a spin-flavor extended WT interaction
\begin{eqnarray}
{\cal L}_{\rm WT}^{\rm sf} &=& \frac{iN}{2f^2}
[\Phi,\partial_\mu\Phi]^j{}_k
\frac{1}{N!}{\cal B}^\dagger_{ji_2\cdots i_N} {\cal B}^{ki_2\cdots i_N} .
\end{eqnarray}
As shown in \cite{GNS05} this extended Lagrangian includes that of SU(3)
in the case $N=N_f=3$, for the $0^-$meson--$1/2^+$baryon 
sector.

Because the covariant derivative acts on each quark index in ${\cal
B}^{i_1\cdots i_N}$, there is an extra factor of $N$ beyond that
coming from the weak coupling constant. This would imply a large $N$
dependence of ${\cal O}(N^0)$ for the WT amplitude, of the same order
as that of the $p$-wave amplitude \cite{DJM94}. This holds for
generic baryons. The situation is more subtle if only baryons with
finite spin and flavor in the large $N$ limit are retained
\cite{GNS05}. In this case the spin-flavor extended WT amplitude is
${\cal O}(N^0)$ only if vector (and/or flavored) mesons are involved,
but ${\cal O}(N^{-1})$ in the pseudoscalar (or flavorless) meson
sector. This is the standard result. Indeed, the
commutator of the absorbed and emitted meson fields in ${\cal L}_{\rm
WT}^{\rm sf}$ just produces a spin-flavor generator to be measured by
the baryonic matrix element. If the mesons carry only flavor but not
spin, the generator will also be purely flavor-like, and flavor was
finite by assumption. This gives a power of $N$ less than for a
generic baryon.

Note the group structure \cite{GNS05}
\begin{equation}
{\cal L}_{\rm WT}^{\rm sf} \sim 
((M^\dagger\otimes M)_{``\treintaycinco_A''}\otimes
(B^\dagger\otimes B)_{``\treintaycinco''})_\uno
\end{equation}
($M$, $M^\dagger$, $B$ and $B^\dagger$ representing annihilation and
creation operators for mesons and baryons) which indicates a purely
antisymmetric adjoint coupling in the $t$-channel. As we will see
below (cf. (\ref{eq:schannel})), in the $s$-channel (or $u$-channel)
there are four possible SU($n$) irreps, and so four independent
couplings for a generic SU($n$) invariant meson-baryon
Hamiltonian.\footnote{This is the generic case. Some irreps may not
exits for particular low values of $n$ or $N$, see (\ref{eq:8}).} In
the $t$-channel one has instead
\begin{eqnarray}
  \hbox{``}\treintaycinco\hbox{''}\otimes\hbox{``}\treintaycinco\hbox{''} &=&
\uno\oplus \hbox{``}\treintaycinco_S\hbox{''}\oplus\hbox{``}\treintaycinco_A\hbox{''}
\oplus\hbox{``}\cuatrocientoscinco\hbox{''}\oplus \cdots
\nonumber \\
  \hbox{``}\cincuentayseis\hbox{''}\otimes\hbox{``}\cincuentayseis^*\hbox{''} &=&
\uno\oplus \hbox{``}\treintaycinco\hbox{''}
\oplus\hbox{``}\cuatrocientoscinco\hbox{''}\oplus \cdots
\end{eqnarray}
where all the irreps not explicited differ from mesons to
baryons. Thus there are again four independent coupling constants in
the $t$-channels, as it should be.
$\hbox{``}\cuatrocientoscinco\hbox{''}$ is the representation with
tableau $[4,2^{n-2}]$. The detailed group structure of WT comes as a
consequence of the SU(n) chiral symmetry requirement \cite{CP76}.

Phenomenological consequences of the extended WT
interaction in the negative parity baryon sector are analyzed in
\cite{GNS05}.

\section{Crossed Projectors}

The coupling of baryons and mesons gives four SU($n$) irreps
\begin{equation}
``\cincuentayseis\hbox{''}\otimes \hbox{``}\treintaycinco\hbox{''}
=
``\cincuentayseis\hbox{''} \oplus ``\setenta\hbox{''} 
\oplus ``\setecientos\hbox{''} \oplus
``\milcientotreintaycuatro\hbox{''}
\label{eq:schannel}
\end{equation}
 with tableaus $[N]$, $[N-1,1]$,
$[N+2,1^{n-2}]$ and $[N+1,2,1^{n-3}]$, respectively. In what follows
they will be called simply irreps 1, 2, 3, and 4. They have dimensions
\begin{eqnarray}
&&
d_1=\left(\matrix{n+N-1 \cr N}\right),\quad
d_2=\frac{(N-1)(n-1)}{n+N-1}\, d_1,
\nonumber \\ &&
d_3= \frac{(n+N-1)(n-1)}{N+1} \,d_1,\quad
\nonumber \\ &&
d_4= \frac{nN(n+N)(n-2)}{(n+N-1)(N+1)} \, d_1 \,.
\label{eq:8}
\end{eqnarray}
For convenience we introduce a vector notation so that $\vec
d=[d_1,d_2,d_3,d_4]$.

A SU($n$) invariant Hamiltonian takes therefore the form (again using
a vector/tensor notation in irrep space)
\begin{equation}
H=\sum_{\alpha=1}^4h_\alpha P_\alpha = {\vec h}\cdot{\vec P}
\end{equation}
where $P_\alpha$ denotes the projector of meson-baryon states on the
irrep $\alpha$. For the extended WT interaction, a direct calculation
using Wick contractions \cite{GNS05} gives
\begin{equation}
\vec\lambda=[n,N+n,-N,1]
\end{equation}
Since $\vec h_{\rm WT}=-\vec\lambda(\sqrt{s}-M)/f^2$, the amplitude
is ${\cal O}(N^0)$ for the irreps 2 and 3 and ${\cal O}(N^{-1})$ for 1
and 4. It is noteworthy the relation
\begin{equation}
{\vec\lambda}\cdot{\vec d}=0
\label{eq:10}
\end{equation}
which embodies the property ${\rm tr}(H_{\rm WT})=0$, in turn coming
from the vanishing of the trace of the meson operator living in the
adjoint representation (obtained from the commutator of the meson
fields).

\begin{figure*}[htb]
\begin{eqnarray}
\tens a &=& \left[ \matrix{ \frac{n(n+N)(N-1)-N^2}{N(n-1)(n+N)} &&
    -\frac{n}{N(n-1)} && \frac{n}{(n-1)(n+N)} &&
    -\frac{n}{N(n-1)(n+N)} \cr -\frac{n(N-1)}{N(n+N-1)} &&
    \frac{n}{N(n+N-1)} && \frac{N-1}{(n+N-1)} && -\frac{N-1}{N(n+N-1)}
    \cr \frac{n(n+N+1)}{(N+1)(n+N)} && \frac{n+N+1}{(N+1)} &&
    \frac{n}{(n+N)(N+1)} && \frac{n+N+1}{(n+N)(N+1)} \cr
    -\frac{n^2(n-2)}{(n-1)(N+1)(n+N-1)} &&
    -\frac{n(n-2)(n+N)}{(n-1)(N+1)(n+N-1)} && \frac{n
      N(n-2)}{(n-1)(N+1)(n+N-1)} &&
    \frac{N(n-1)(n+N)+1}{(n-1)(N+1)(n+N-1)} \cr } \right]
\label{eq:tabla}
\end{eqnarray}
\end{figure*}

As said the WT amplitude is purely $\hbox{``}\treintaycinco\hbox{''}$
in the $t$-channel and $\vec\lambda$ reflects this group structure as
seen in the $s$-channel. In what follows we investigate how the
various irrep in the $t$- and $u$-channels are seen in the
$s$-channel. The latter channel corresponds to crossing of the meson
legs. For instance a direct baryon-pole ($s$-channel) mechanism would
give a pure \hbox{``}\cincuentayseis\hbox{''} coupling\footnote{We
assume here an hypothetical SU($n$) invariant meson-baryon-baryon
coupling. The physical coupling is $p$-wave, hence the orbital
momentum would have to be included.}, i.e., $\vec h_{\rm
dbp}\propto[1,0,0,0]$, while a crossed baryon-pole ($u$-channel)
mechanism
\begin{equation}
((M\otimes B^\dagger)_{``\cincuentayseis''{}^*}\otimes 
(M^\dagger\otimes B)_{``\cincuentayseis''}
)_\uno
\end{equation}
gives instead (see Appendix D of [5$a$])
\begin{equation}
\vec\lambda_{\rm cbp}=\left[N+n-1-\frac{N}{n}-\frac{n}{N},
-1-\frac{n}{N},1,-\frac{1}{N}\right]
\end{equation}
(the normalization is arbitrary). While the effect of crossing for
other irreps could be computed as in \cite{GNS05}, by using Wick
contractions, we present here a shortcut. To this end we introduce the
crossed projectors
\begin{equation}
\langle m_1, b_1|P^c_\alpha|m_2,b_2\rangle
= \langle \overline{m_2}, b_1|P_\alpha|\overline{m_1},b_2\rangle
\end{equation}
where $|\overline{m}\rangle=C|m\rangle$ indicates the crossed meson
state. Since the $P^c_\alpha$ themselves are invariant operators, we
have
\begin{equation}
P^c_\alpha=\sum_\beta a_{\alpha\beta}P_\beta
,\quad\hbox{or}
\quad
{\vec P}^c={\tens a}\cdot {\vec P}
\end{equation}
where the (real) crossing coefficients $a_{\alpha\beta}$ are those to
be determined. They describe the $u$-channel irreps as seen in the
$s$-channel.

For the group SU(2) the (antilinear) operator $C$ is just
$|\overline{j,m}\rangle=(-1)^{j-m}|j,-m\rangle$, however, in the
general case, the only property needed of $C$ is to be a unitary
conjugation, i.e., $CC^\dagger=C^2=1$. Together with general projector
properties
\begin{eqnarray}
&& P_\alpha=P_\alpha^\dagger, \quad
P_\alpha P_\beta=\delta_{\alpha\beta}P_\alpha \,,\quad
\nonumber \\
&& 
{\rm tr}\, P_\alpha=d_\alpha
\,,\quad \sum_\alpha P_\alpha =1 \,,
\end{eqnarray}
one easily establishes the relations
\begin{eqnarray}
&& \sum_\gamma
a_{\alpha\gamma}a_{\gamma\beta}=\delta_{\alpha\beta},\quad 
\sum_\gamma a_{\alpha\gamma}a_{\beta\gamma}\,d_\gamma
=d_\alpha \delta_{\alpha\beta},
 \nonumber \\ &&
\sum_\alpha
a_{\alpha\beta}=1, \quad 
\sum_\beta a_{\alpha\beta}\,d_\beta=d_\alpha .
\label{eq:11}
\end{eqnarray}
The first two relations imply that the matrix 
\begin{equation}{\hat a}_{\alpha\beta}
=d_\alpha^{-1/2}a_{\alpha\beta}\,d_\beta^{1/2}
\end{equation}
satisfies $\hat{\tens a}=\hat{\tens a}^T=\hat{\tens a}^{-1}$, i.e., it
has four orthonormal eigenvectors ${\hat e_\lambda}$ with eigenvalues
$\pm 1$.  As a consequence $\hat{\tens a}$ can be written as
$P_{+1}-P_{-1}=1-2P_{-1}$ ($P_{\pm 1}$ being the orthogonal projectors
on the $\pm 1$ subspaces of the four dimensional irrep space) and
$\tens a$ is just $\hat{\tens a}$ expressed in a non-orthonormal
basis. The third and fourth relations in (\ref{eq:11}) identify $[1,1,1,1]$ as one such
left eigenvector of $\tens a$, with eigenvalue $+1$, and $\vec d$ as
its right eigenvector version.

The left eigenvector $\vec\lambda_{\uno}=[1,1,1,1]$ can also be identified as the
$s$-channel view of the $t$-channel coupling $((M^\dagger\otimes
M)_\uno\otimes (B^\dagger\otimes B)_\uno )_\uno$ (this operator just
counts the number of mesons and baryons), and the eigenvalue  $+1$
indicates that $\uno$ comes from a symmetric coupling of
$\hbox{``}\treintaycinco\hbox{''}\otimes\hbox{``}\treintaycinco\hbox{''}$.

On the other hand, the WT interaction (irrep
$\hbox{``}\treintaycinco_A\hbox{''}$ in the $t$-channel) is
antisymmetric under meson crossing and so $\vec\lambda$ is a left
eigenvector of $\tens a$ with eigenvalue $-1$ (correspondingly, it is
orthogonal to $\vec\lambda_1$, see (\ref{eq:10}).) The two other left
eigenvectors correspond to $\hbox{``}\treintaycinco_S\hbox{''}$ and
$\hbox{``}\cuatrocientoscinco\hbox{''}$ in the $t$-channel. Since they
are both symmetric under crossing, they carry eigenvalue $+1$.

In view of the fact that $\vec\lambda_{``\treintaycinco_A''}
=\vec\lambda$ is the unique eigenvector with eigenvalue $-1$, we can
directly reconstruct $\tens a$ as follows\footnote{Since $\hat{\tens
a}=1-2P_{-1}= 1-2\,{\hat e}_{-1}\otimes{\hat e}_{-1}$.}
\begin{equation}
a_{\alpha\beta}=\delta_{\alpha\beta}
-2\frac{\lambda_\alpha\lambda_\beta\, d_\alpha}{\sum_\gamma \lambda^2_\gamma \,d_\gamma} \,.
\end{equation}
The matrix is displayed in (\ref{eq:tabla}).
Note that $\vec\lambda_{\rm cbp}$ is correctly reproduced (up to a
factor) by the first row of $\tens a$. This matrix contains all
$u$-channel results. For the ``$\treintaycinco_S$'' $t$-channel, a direct
Wick computation gives
\begin{equation}
\vec\lambda_{``\treintaycinco_S''}=
[(\frac{2}{n}+\frac{1}{N})(n-2),
(\frac{1}{n}+\frac{1}{N})(n-2),
\frac{n-2}{n},
-\frac{1}{N}-\frac{2}{n}
]
\end{equation}
As can be verified it is a $+1$ left eigenvector of $\tens a$ and is
orthogonal to $\vec\lambda_\uno$ and
$\vec\lambda_{``\treintaycinco_A''}$. Finally, by orthogonality we
obtain
\begin{equation}
\vec\lambda_{``\cuatrocientoscinco''}= \left[
n,
-\frac{n+N}{N-1},
-\frac{N}{n+N+1},
\frac{1}{n}
\right] \,.
\end{equation}


\begin{acknowledgement}
This research was supported by DGI and FEDER funds, under contract
FIS2005-00810, by the Junta de Andaluc{\'\i}a (FQM-225) and the EU network
EURIDICE, under contract HPRN-CT-2002-00311. It is also part of the EU
integrated infrastructure initiative Hadron Physics Project under
contract number RII3-CT-2004-506078.
\end{acknowledgement}


\end{document}